\documentclass[twocolumn,showpacs,preprintnumbers,amsmath,amssymb]{revtex4}
\usepackage{graphicx}
\usepackage{dcolumn}
\usepackage{bm}
\begin{document}
\title{Diquark and light four-quark states}

\author{Ailin Zhang$^1$, Tao Huang$^2$ and Tom G.\ Steele$^3$}
\affiliation{ $^1$Department of Physics, Shanghai University,
Shanghai, 200444, China;\\
$^2$Institute of High Energy Physics, Chinese Academy of Science, P.
O. Box 918 (4), Beijing, 100049,
China;\\
$^3$ Department of Physics and Engineering Physics, University of
Saskatchewan, Saskatoon, SK, S7N 5E2, Canada}


\begin{abstract}
Four-quark states with different internal clusters are discussed
within the constituent quark model. It is pointed out that the
diquark concept is not meaningful in the construction of a
tetraquark interpolating current in the QCD sum rule approach, and
hence existing sum-rule studies of four-quark states  are
incomplete. An updated QCD sum-rule determination of the properties
of diquark clusters is then used as input for the constituent quark
model to obtain the masses of light $0^{++}$ tetraquark states ({\it
i.e.\ }a bound state of two diquark clusters). The results support
the identification of $\sigma(600)$, $f_0(980)$ and $a_0(980)$ as
the $0^{++}$ light tetraquark states, and seem to be inconsistent
with the tetraquark state interpretation of the new BES observations
of the near-threshold $p\bar p$ enhancements, $X(1835)$ and
$X(1812)$, with the possible exception that $X(1576)$ may be an
``exotic" first orbital excitation of $f_0(980)$ or $a_0(980)$.
\end{abstract}

 \maketitle

\section{Introduction}
In the quark model, a meson consists of a quark and an antiquark,
while a baryon consists of three quarks. Exotic hadrons are those
beyond these naive $q\bar q$ mesons and $qqq$ baryons. Among the
exotic hadrons, four-quark state scenarios are of both theoretical
and experimental interest. A four-quark state was predicted to exist
early in a consistent description of the hadron scattering
amplitudes \cite{four1}, and their properties have been studied in
many models
\cite{four2,chan,four3,mole1,mole2,mole3,isgur2,torn,frank,wong,maiani,maiani1,fariborz,others}.

Quarks/anti-quarks in four-quark states may have complicated
correlations and form different clusters. According to the spatial
extension of the clusters, four-quark states may be composed of a
diquark and an anti-diquark \cite{four1,four2,four3}, or be composed
of two $q\bar q$ clusters (including the lightly bounded ``molecular
states") \cite{mole1,mole2,mole3,isgur2,wong}. It could be expected
that the properties of these two kinds of four-quark states are
different because of their different internal structures though the
dynamics among quarks in four-quark state is still unclear.

So far, no four-quark state has been confirmed experimentally.
However, there are some four-quark candidates in experiments,
including $f_0(600)$ (or $\sigma$), $f_0(980)$, $a_0(980)$ and the
unconfirmed $\kappa(800)$. They have a long history of being
interpreted as four-quark states \cite{four2,pdg}. Recently,
$X(3872)$ has been observed by Belle \cite{belle} in exclusive B
decays and $Y(4260)$ has been observed by BaBar \cite{babar} in
initial-state radiation events. These newly observed states have
also been interpreted as four-quark states in some literatures
\cite{x1,x2,wong,x3,maiani,x4,maiani1}.

In particular, the BES collaboration has reported some new
observations in low energy region. Apart from the near-threshold
$p\bar p$ enhancement \cite{bes1}, $X(1835)$ was observed by BES
\cite{bes2} in the decay
\begin{eqnarray*}
J/\Psi\to\gamma\pi^+\pi^-\eta^\prime,
\end{eqnarray*}
with $M=1833.7\pm 6.1(stat)\pm 2.7(syst)$ MeV, $\Gamma=67.7\pm
20.3\pm 7.7$ MeV. $X(1812)$ was observed by BES \cite{bes3} in the
doubly OZI-suppressed decay
\begin{eqnarray*}
J/\Psi\to\gamma\omega\phi
\end{eqnarray*}
with $M=1812^{+19}_{-26}(stat)\pm 18(syst)$ MeV, $\Gamma=105\pm
20\pm 28$ MeV. Most recently, $X(1576)$ was observed by BES at the
$K^+K^-$ invariant mass in the decay \cite{x1576}
\begin{eqnarray*}
J/\Psi\to K^+K^-\pi^0
\end{eqnarray*}
with pole position $1576^{+49}_{-55}(stat)^{+98}_{-91}(syst)$
MeV-$i(409^{+11}_{-12}(stat)^{+32}_{-67})$ MeV. The $J^{PC}$ of this
broad peak is $1^{--}$. Whether these observations by BES will be
confirmed or not by other experimental groups in the future,
four-quark candidates interpretations \cite{bes4,bes5,shen,lipkin}
to them have appeared.

Diquarks are relevant to the understanding of four-quark states in
constituent quark models. Diquarks were first mentioned by Gell-Mann
\cite{gell} and have been applied successfully to many
strong-interaction phenomena \cite{di1,di2,di3}. Interest in
diquarks has been revived by a variety of new experimental
observations \cite{di4,di5,di6,di7}. As pointed out in references
\cite{di7,maiani,di9},  diquark clusters in hadrons are in fact a
kind of strong correlation between pairs of quarks; this diquark
correlation may be most important for the light multi-quark states.

Motivated by these theoretical and experimental developments, we
construct light four-quark states via diquark clusters. The diquarks
are regarded as hadronic constituents, with their masses
 determined via QCD sum rules. As an approximation, the
masses determined in this way may be regarded as the constituent
masses of diquarks. Accordingly, masses of these light tetraquark
states are then obtained within the constituent quark models. As
outlined below, this approach obviates the difficulties in a
sum-rule approach based purely on local interpolating currents for
four-quark/tetraquark states.
 Based
on these results, the tetraquark state possibility of the new
observations by BES is analyzed. We emphasize that the existence of
diquark clusters within exotic states is an open question.  Our work
explores the implications of diquarks for the mass spectrum of the
tetraquark states.

\section{Four-quark states and diquark clusters}
As mentioned above, four-quark states have complex internal color,
flavor and spin structure. First, we consider the color, flavor and
spin correlation of the $(\bar q\bar q)(qq)$ four-quark state
(tetraquark) and the $(\bar qq)(\bar qq)$ four-quark state.

In the $(\bar q\bar q)(qq)$ configuration , the quark $q$ is in the
color fundamental representation $3$, while the anti-quark $\bar q$
is in the color representation $\bar 3$. Two quarks give $3\otimes
3=6\oplus\bar 3$, and two anti-quarks give $\bar 3\otimes \bar
3=\bar 6\oplus 3$. Therefore, the final color singlet four-quark
state may be produced from $\bar 3\otimes 3$ or $\bar 6\times 6$.

In the $(\bar qq)(\bar qq)$ configuration, a quark and an anti-quark
give the color representations $3\otimes \bar 3=8\oplus 1$, another
quark anti-quark pairs give the same result. The final color singlet
are therefore produced from the $1\otimes 1$ or $8\times 8$. The
former $1\otimes 1$ four-quark state is usually called a
``molecule".

It is difficult to distinguish between these internal color
configurations because the final state is a color singlet. For the
same reason, it is difficult to distinguish between four-quark and
normal $q\bar q$ mesons.  An understanding of the internal dynamics
of the quarks in the hadrons would be necessary to distinguish
between these various configurations.

In the flavor $SU(3)$ approximation, both the $(\bar q\bar q)(qq)$
and $(\bar qq)(\bar qq)$ configurations give the same flavor
multiplets: $(3\otimes 3)\otimes (\bar 3\otimes \bar 3)=(3\otimes
\bar 3)\otimes (3\otimes \bar 3)=27\oplus 10\oplus \bar {10}\oplus
8\oplus 8\oplus 8\oplus 8\oplus 1\oplus 1$. Obviously, the flavor
structure of $(qq)(\bar q\bar q)$ and $(\bar qq)(\bar qq)$
four-quarks cannot be distinguished. Except for the flavor octet or
singlet (crypto-exotic), other flavor structures in the four-quark
state do not exist in normal $q\bar q$ meson. The crypto-exotic
four-quark states will mix with normal $q\bar q$ mesons, while
four-quark state may exhibit its exotic flavor explicitly in a
different way as that in the normal $q\bar q$ meson. Accordingly,
experiments could be designed to detect four-quark states due to
their exotic flavor structures. The spin in four-quark state will
couple (correlate) and give complicated representations as those
mentioned in reference \cite{four2}.

In fact, the correlation of color, flavor and spin in hadron is
inter-related since their total correlation has to obey some
symmetry constraints. In the $(\bar q\bar q)(qq)$ configuration,
according to references \cite{di4,di7,di9}, the two quarks correlate
antisymmetrically in color, flavor and spin, separately, and thereby
attract one another forming a ``good" diquark cluster. In other
words, the two quarks are most possible in the color, flavor and
spin representations $\bar 3$, $\bar 3$ and $0$ and form the ``good"
diquark. This antisymmetric diquark cluster will be denoted as
$[qq]$ in the following.

The composition of light tetraquark states with different isospin
are $[\bar q\bar q][qq]$, $[\bar s\bar q][qq]$, $[\bar s\bar
q][sq]$, $\ldots$ as in references \cite{maiani,maiani2} with
$q=u,~d$. The orbital excitation between the diquark and
anti-diquark may make different kinds of four-quark states.

\section{Diquark four-quark states and new observations by BES}
QCD sum rules \cite{sum} is believed to be a good tool to study
hadron physics and nonperturbative QCD interactions. With this
approach, four-quark states have been investigated. In the
investigations, different interpolating currents (operators) have
been used. {\it e.g.,} $(\bar qq)(\bar qq)$ currents were used in
reference \cite{zhang},  $(\bar q\bar q)(qq)$ diquark anti-diquark
currents were employed in references \cite{nielsen,kim,zhu}, and
both currents were studied in reference \cite{schafer}. These
investigations are interesting and may give some hints to our
understanding of four-quark states. However, some of the conclusions
related to  the diquark concept based on these studies are not
definitive. Diquark may be the reality in constituent quark models,
but a diquark interpretation is not meaningful in the framework of
QCD sum rules.

As well known, each hadron is in color singlet. Internal color
configurations in $(\bar qq)(\bar qq)$ and $(\bar q\bar q)(qq)$
could not be detected directly through those interpolating currents
except through  special observable which may detect those different
couplings of currents. Unfortunately, no such an observable has yet
been formulated. Moreover, these color configurations can not be
distinguished from the normal $\bar qq$ mesons either, which implies
that the mixing between four-quark state and normal mesons may be
crucial.

From the analyses in last section, internal constituent flavor
configurations in $(\bar qq)(\bar qq)$ and $(\bar q\bar q)(qq)$ give
the same representations. Their internal flavor configurations
therefore could not be detected directly through those interpolating
currents either except for  flavor exotic currents.

This point is much more easy to be realized in other ways. On one
hand, the Fierz transformation will turn the $(\bar qq)(\bar qq)$
current into the $(\bar q\bar q)(qq)$ current and vice-versa. On the
other hand, these two kinds of currents can mix with each other
under renormalization. Therefore, it may not be meaningful to talk
about diquarks in the framework of sum rules based on local
interpolating currents. In order to construct precise four-quark
state sum rules, both $(\bar qq)(\bar qq)$ and $(\bar q\bar q)(qq)$
currents have to be used. Similar situations occurred in the study
of baryons with sum rules \cite{baryon1,baryon2}.  We emphasize that
the inability to distinguish between the local $(\bar qq)(\bar qq)$
interpolating currents $(\bar q\bar q)(qq)$ in the QCD sum-rule
context does not occur, for example, in potential models
\cite{brink}.

In principle, there is no direct way to turn the operator picture
into the constituent quark picture \cite{melikhov}, and the internal
constituent quark structures can not be detected directly through
the couplings of local interpolating currents to hadrons
\footnote{Ailin Zhang is grateful to Prof.\ M.\ Shifman for the
helpful communications on this viewpoint}.

To make the sum rule analyses reliable and predictable, it is better
to use flavor exotic currents. For normal interpolating currents,
the mixing has to be incorporated in the sum rule. For example, a
renormalization invariant mixed current may be required, and the
saturation of the spectral density with mixed hadrons should be
taken into account. To avoid such complexities, we will not study
the four-quark state in terms of four-quark currents in this
article. Instead, we will determine the mass of two-quark cluster
via QCD sum rules, and subsequently construct the four-quark state
in terms of these diquark constituents.

From a practical viewpoint, an analysis of tetraquark states that
employs QCD sum-rules for the diquark clusters to obtain inputs for
the constituent quark model is likely the only feasible approach to
the study of tetraquark states. Correlators of tetraquark/four-quark
interpolating currents have only been calculated to leading order in
$\alpha_s$ \cite{zhang,nielsen,kim,zhu,schafer}. Obtaining the
$\alpha_s$ corrections that would be needed to establish a reliable
QCD sum-rule prediction would involve four-loop calculations and the
renormalization of the dimension-six composite operators. Although
the loop calculations are feasible in the chiral limit, the
renormalization of these operators is unknown beyond a single quark
flavour \cite{tarrach} and extensions to more complicated flavour
structures would be exceptionally difficult, particularly with heavy
flavours.

Lattice investigations pointed out that the force between colored
clusters is universal \cite{ukawa,bali}. Since the diquark and the
anti-diquark clusters in $(\bar q\bar q)(qq)$ tetraquark state have
the same color representations as $\bar q$ and q in normal meson
($3\otimes \bar 3\rightarrow 1$), they are expected to have similar
strong dynamics as the normal constituent quarks. It is thus
reasonable to expect that these diquark and anti-diquark clusters
will form a tetraquark state just as the constituent quark and
anti-quark construct a normal meson. A four-quark state with
constituent configuration $(\bar qq)(\bar qq)$ has no correspondence
to existing hadrons, and will not be considered in our analysis.

Though the diquark is not an isolated cluster in hadron, it may be
approximately regarded a bound state composed of two quarks and may
be used as degree of freedom. Historically, the effective diquark
masses and couplings were derived with QCD sum rules approach in the
study of weak decays in references \cite{di2,di10} .

Before proceeding, we give a simple description  for the dynamics in
tetraquark state in our model. The diquark cluster is regarded as a
constituent similar to a constituent quark. The strong dynamics
between the diquark and the anti-diquark in tetraquark states is
supposed to be similar to that between the quark and the anti-quark
in mesons \cite{ukawa,bali}. As an approximation, the masses of
diquarks obtained with sum rule approach are regarded as the
constituent masses in constituent quark model. Therefore the
spectrum of the four-quark states could be obtained  as in
references \cite{maiani1,maiani3}.

Since the diquark and anti-diquark are supposed to be in the $0^+$
``good" configuration, the P-parity and the C-parity of the neutral
tetraquark states are the same $(-1)^L$, where $L$ is the orbital
angular momentum between the diquark and the anti-diquark.
Accordingly, possible $J^{PC}$ of these kinds of tetraquark states
are \cite{zhang1} $0^{++}~(L=0)$, $1^{--}~(L=1)$, $2^{++}~(L=2)$,
$\ldots$. Thus the mass of the $0^{++}~(L=0)$ tetraquark is
\cite{maiani}
\begin{equation}
M_{4q}=m_d+m_{\bar d}-3(\kappa_{qq})_{\bar 3}, \label{eq1}
\end{equation}
where $m_d=m_{qq}$, $m_{\bar d}=m_{\bar q\bar q}$ are the
constituent masses of diquark and anti-diquark, respectively.
Similarly, the mass of the $1^{--}~(L=1)$ and the $2^{++}~(L=2)$
tetraquark is
\begin{equation}
M_{4q}=m_d+m_{\bar d}+B_{d\bar d}{L(L+1)\over 2}, \label{eq2}
\end{equation}
where $B_{d\bar d}=B^\prime_q,~B^\prime_{1s},~B^\prime_{2s}$.
$B^\prime_q> B^\prime_{1s}>B^\prime_{2s}$ \cite{maiani1,maiani3},
and they denote the coefficients with zero, one and two strange
quarks, respectively. If $B_{d\bar d}\propto \alpha^2_s M$, they are
very sensitive to $\Lambda_{QCD}$.

In references \cite{di2,di10}, the flavor combinations $(qq)
(q=u,d)$, $(sq)$ and $(ss)$ were used to construct different flavor
contents of the diquark to simplify  calculation. The flavor $(sq)$
diquark current was taken to be \cite{di2,di10}
\begin{eqnarray}
j_i(x)=\epsilon_{ijk}s^T_j(x)COq_k(x),
\end{eqnarray}
where $i,~j,~k$ are color indices, $C$ is the charge conjugation
matrix, and $O=\gamma_5,~1,~\gamma_\mu,~\gamma_\mu\gamma_5$ are the
Lorentz structures corresponding to quantum numbers
$J^{P}=0^+,~0^-,~1^+$, and $1^-$, respectively.

After constructing a gauge invariant correlator, theoretical
expressions of $\Pi(Q^2=-q^2)$ for $(sq)_i$ were computed in
references \cite{di2,di10}
\begin{eqnarray}
\Pi(Q^2)={3\over 4\pi^2}(1+2{m^2_s\over Q^2}+{17\over
6}{\alpha_s\over \pi}-{1\over 2}{\alpha_s\over \pi}\ln{Q^2\over
\mu^2})Q^2\ln{Q^2\over \mu^2}\\\nonumber -(2m_s-m_q){\langle \bar
qq\rangle\over Q^2}-(2m_q-m_s){\langle \bar ss\rangle\over
Q^2}+{1\over 8}\langle {\alpha_s\over \pi}G^2\rangle{1\over
Q^2}\\\nonumber +8\pi\kappa\alpha_s{\langle \bar
qq\rangle\langle\bar ss\rangle\over Q^4}-{16\pi\over
27}\kappa\alpha_s{\langle \bar qq\rangle^2+\langle\bar
ss\rangle^2\over Q^4},
\end{eqnarray}
where $\langle\bar qq\rangle$, $\langle\bar ss\rangle$, $\langle
{\alpha_s\over\pi} G^2\rangle$ are two quark and two gluon
condensates, and $\kappa$ denotes the deviations from factorization
approximation of four quark condensates($\kappa=1$ for ideal
factorization).

In this paper, we perform an updated sum-rule study of the $J^P=0^+$
``good" diquark. Within the single pole and continuum approximation,
the mass of diquark is able to be obtained from
$m^2_{0^+}=-{\partial\over
\partial\tau}\ln R_k(\tau,s_0)$ with
\begin{eqnarray}
R_k(\tau,
s_0)&=&\frac{1}{\tau}\hat{L}[(-Q^2)^k\{\Pi(Q^2)-\sum\limits_{k=0}^{n-1}a_k(-Q^2)^k\}]\\\nonumber
&-&\frac{1}{\pi}\int_{s_0}^{+\infty}s^k
e^{-s\tau}Im\Pi^{\{pert\}}(s)ds\\\nonumber
&=&\frac{1}{\pi}\int_{0}^{s_0}s^k e^{-s\tau}Im\Pi(s)d s.
\end{eqnarray}

$R_0(\tau,s_0)$ for $(sq)_i$ with renormalization group improvement
follows from references \cite{di2,di10}
\begin{eqnarray}
R_0&=&({\alpha_s(\mu^2)\over \alpha_s(1/\tau)})^{-4/9}{3\over
4\pi^2}{1\over \tau^2}\times\\\nonumber &&\{(1+{17\over
6}{\alpha_s(1/\tau)\over
\pi})[1-(1+{s_0\tau})e^{-s_0\tau}]-\\\nonumber
&&(1-\gamma_E){\alpha_s(1/\tau)\over\pi}\Phi(s_0\tau)-2m^2_s\tau
(1-e^{-s_0\tau})\}-\\\nonumber &&[(2m_s-m_q)\langle \bar
uu\rangle+(2m_q-m_s)\langle \bar ss\rangle]+{1\over 8}\langle
{\alpha_s\over\pi}G^2\rangle\\\nonumber
&&+[8\pi\kappa\alpha_s\langle \bar uu\rangle \langle \bar
ss\rangle-{16\pi\over 27}\kappa\alpha_s(\langle \bar
uu\rangle^2+\langle \bar ss\rangle^2)]\tau,
\end{eqnarray}
respectively, where $\gamma_E$ is the Euler constant, and $\Phi(x)$
was given in reference \cite{di10}. The result of $(qq)$ diquark
could be obtained by the replacement of s quark with q quark in
previous formulas.

To obtain an updated numerical result, the input parameters are
chosen as those in references \cite{para1,para2,pdg}:
$m_u=m_d=m_q=5~MeV$, $m_s=120~MeV$, $\langle \bar qq
\rangle=-(260~MeV)^3$, $\langle \bar ss \rangle =0.8\langle \bar qq
\rangle$, $\langle {\alpha_s\over\pi}GG \rangle=0.012~GeV^4$,
$\kappa\alpha_s\langle \bar qq \rangle^2=5.8\times 10^{-4}~GeV^6$,
$\Lambda=375~MeV$, $\alpha_s(\mu^2)=4\pi/ 9\ln {\mu^2\over
\Lambda^2}$ and renormalization scale $\mu\sim 1$ GeV.

The sum rules for $(qq)$ and $(sq)$ diquarks are good. $m^2_{0^+}$
increases slowly but monotonically with the increase of $s_0$. To
obtain a reliable prediction, the continuum $s_0$ is chosen as low
as possible, and to make the Borel window as wide as possible in the
case of keeping $R_k$ positive. The variation of $m^2_{0^+}$ for
$(qq)_i$ and $(sq)_i$ with Borel variable $\tau$ are respectively
shown in Figure 1 and Figure 2. The variation of $m^2_{0^+}$ for
$(qq)_i$ and $(sq)_i$ with $s_0$ are shown in Figure 3 and Figure 4.
The most suitable $m_{qq}$ and $m_{sq}$ are found to be around $400$
MeV and $460$ MeV, respectively. They are roughly comparable in
scale to the constituent quark. The results obtained here are
consistent with the fit of Maiani {\it et al.} \cite{maiani1}, where
the $m_{[ud]}=395$ MeV and $m_{[sq]}=590$ MeV.
\begin {figure}
\includegraphics[scale=0.8]{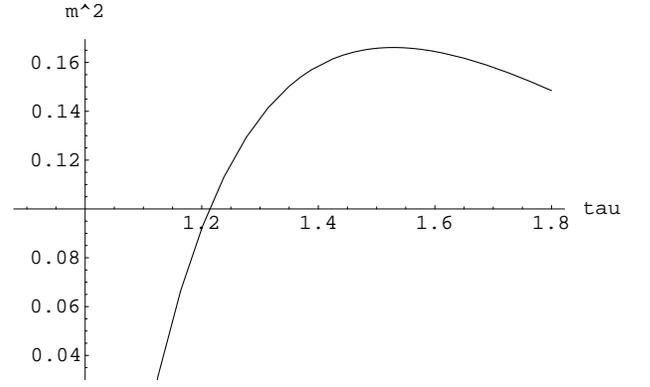}
\caption{$\tau$ dependence of $m^2_{0^+}$ for $qq$ good diquark with
$s_0=1.2$ GeV$^2$.}
\end {figure}

\begin {figure}
\includegraphics[scale=0.8]{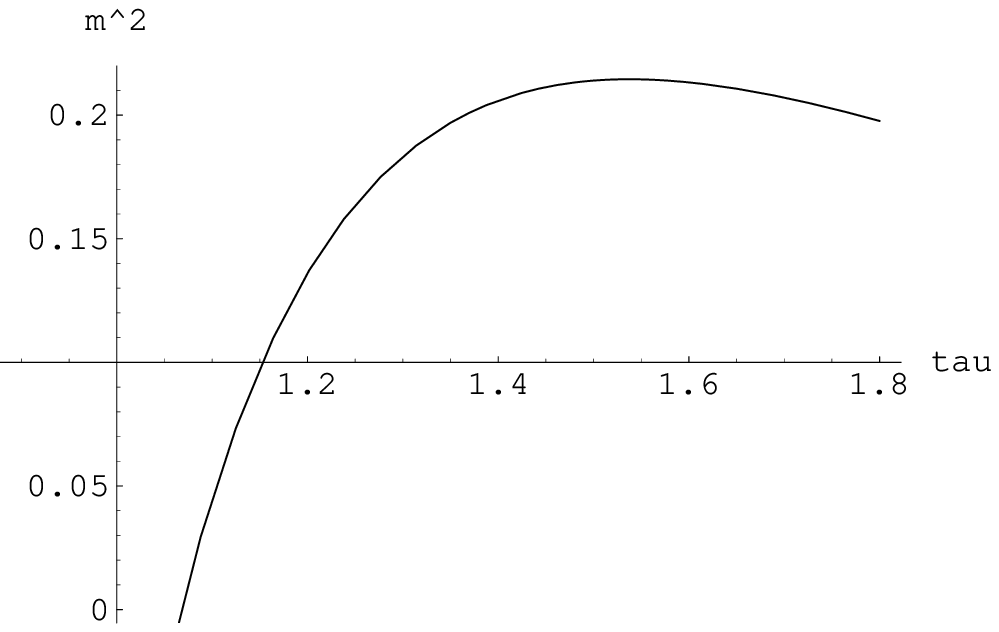}
\caption{$\tau$ dependence of $m^2_{0^+}$ for $sq$ good diquark with
$s_0=1.2$ GeV$^2$.}
\end {figure}

\begin {figure}
\includegraphics[scale=0.8]{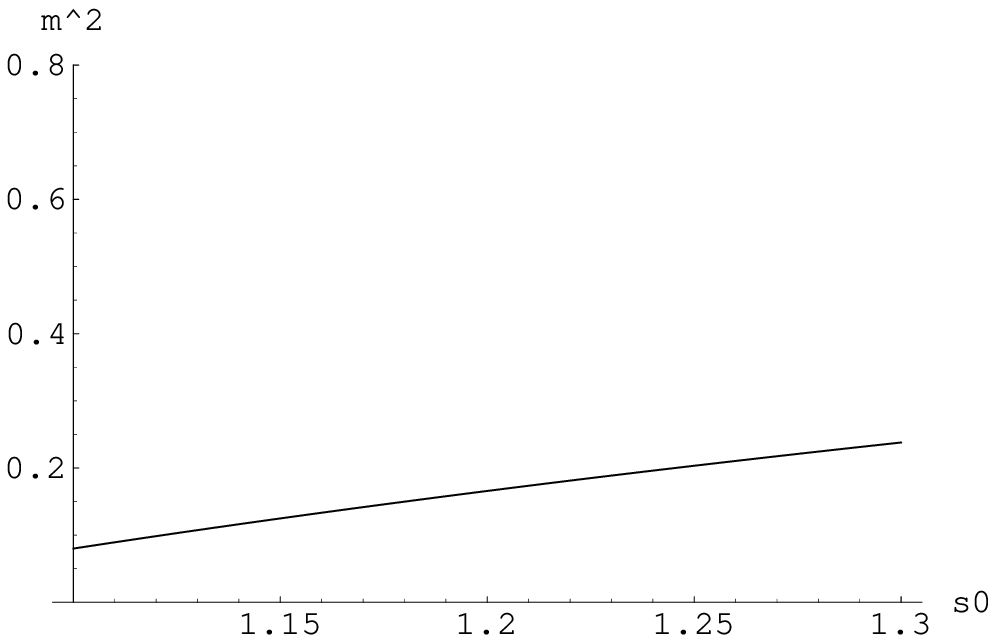}
\caption{$s_0$ dependence of $m^2_{0^+}$ for $qq$ good diquark with
$\tau=1.5$ GeV$^{-2}$.}
\end {figure}

\begin {figure}
\includegraphics[scale=0.8]{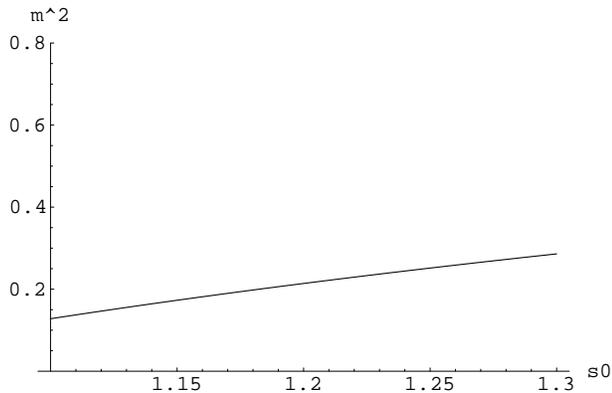}
\caption{$s_0$ dependence of $m^2_{0^+}$ for $sq$ good diquark with
$\tau=1.5$ GeV$^{-2}$.}
\end {figure}

Now that the properties of  the diquark cluster are determined, it
is possible to construct a four-quark provided that the quark
dynamics in hadrons is known. There exist many works which address
internal quark dynamics
\cite{four2,chan,mole3,model1,tye,isgur,eichten,glozman,metsch,ebert}.
Unfortunately, there is no one approach directly from QCD---a very
difficult task. In this paper, the prescription in reference
\cite{model1,mole3,maiani1} is used. Hadron masses were obtained
from constituent quarks and their spin-dependent interactions
\begin{eqnarray}
H=\sum_i m_i+\sum_{i<j}2 \kappa_{ij}(S_i\cdot S_j),
\end{eqnarray}
where the $\kappa_{ij}$ are coefficients, and the sum runs over all
the constituents.  The resulting masses for four-quark states formed
via diquark clusters  are given in Eqs.~(\ref{eq1}) and (\ref{eq2}).

In terms of the $(\kappa_{qq})_{\bar 3}=103$ MeV and
$(\kappa_{sq})_{\bar 3}=64$ MeV \cite{maiani1}, the masses of
$0^{++}$ tetraquark states $[\bar q\bar q][qq]$, $[\bar q\bar
q][sq]$ ($[\bar s\bar q][qq]$) and $[\bar s\bar q][sq]$ are found
via (\ref{eq1}) to be $\sim 490$ MeV, $\sim 610$ MeV and $\sim 730$
MeV, respectively.

As a more ``accurate" prescription, the masses of tetraquark states
could be obtained with constituent diquarks (with masses obtained by
sum rules) in the coulombic and linear confinement potentials, an
approach which is beyond our scope of this article. Taking into
account the decay features, in the approximation we used, it is
reasonable to identify $f_0(600)$ (or $\sigma$), $f_0(980)$,
$a_0(980)$ and the unconfirmed $\kappa(800)$ as the $0^{++}$ light
tetraquark states.  This identification agrees with the conclusions
of analyses based on chiral Lagrangians \cite{fariborz} and chiral
perturbation theory  \cite{pelaez}.

Masses of the $1^{--}$ orbital excited $[\bar q\bar q][qq]$, $[\bar
s\bar q][qq]$ and $[\bar s\bar q][sq]$ are respectively determined
via (\ref{eq2}):
\begin{eqnarray*}
\sim 490+B^\prime_q~~MeV,\\\nonumber \sim
610+B^\prime_{1s}~~MeV,\\\nonumber \sim 730+B^\prime_{2s}~~MeV.
\end{eqnarray*}
$B_{d\bar d}$ are very sensitive to $\Lambda_{QCD}$. It is not
reliable to obtain the masses of orbital excited tetraquark states
with them. Therefore theoretical estimates of the masses of these
tetraquark states are not given here.  However, if a tetraquark
state is confirmed in the future, some constraints on these
$B_{d\bar d}$ could then be derived.

Now we turn our attention to the tetraquark state possibility of the
new observations by BES mentioned above. The $p\bar p$ threshold
enhancement was also seen by Belle \cite{belle2}, while other
observations have not been seen by any other experimental group. The
P-parity of $p\bar p$ is likely to be $C=+1$, and its quantum number
assignment is consistent with either $J^{PC}=0^{-+}$ or $0^{++}$.
$X(1835)$ is found consistent with expectations for the state that
produces the strong $p\bar p$ mass threshold enhancement. $X(1812)$
favors $J^P=0^+$.

These three signals are very unlikely to be the tetraquark states.
As analyzed above, if they have definite $C$ parity, possible
$J^{PC}$ for tetraquark state interpretation is constrained. These
$0^{++}$ or $0^{-+}$ observations lie above the predicted $0^{++}$
or $0^{-+}$ light tetraquark states (Assuming that the mass
difference between the $0^{++}$ and the $0^{-+}$ is not very large).
A $0^{-+}$ tetraquark state requires the bad diquark.

It was argued that the light (orbital angular momentum between the
diquark and the anti-diquark $L=0$) $q^2\bar q^2$ states may decay
into meson-meson channels, while the heavier ones ($L\geq 1$) decay
mainly into baryon-antibaryon channels \cite{four2}, and may decay
into another light tetraquark state ($L=0$) \cite{chan}. $X(1576)$
has large decay width, and its isospin has not been determined.
$X(1576)$ behaves unlike a normal meson and may be a $1^{--}$
tetraquark state \cite{shen,lipkin,ding}. If $X(1576)$ is a $1^{--}$
four-quark state, it is likely to be the $(\bar s\bar q)(sq)$
orbital excited tetraquark state. It may be the first orbital
excitation of $a_0(980)$ if its isospin $I=1$, and it may be the
first orbital excitation of $f_0(980)$ if its isospin $I=0$.

If this suggestion is true, the $B^\prime_{2s}\sim 586$ MeV. This
$B^\prime_{2s}$ is large compared to those for normal mesons, which
may imply an ``exotic" orbital excitation. In the meantime, other
$1^{--}$ orbital excited tetraquark states corresponding to $(\bar
q\bar q)(qq)$ ($\sim 1400$ MeV) and $(\bar s\bar q)(qq)$ ($\sim
1500$ MeV) are expected. Both of them may have broad widths. In this
case, their radiative decays into the scalars would be explicitly
indications.

\section{Conclusions and discussions}

The intrinsic color, flavor and spin configurations of two different
kinds of spatially-extended  four-quark states are discussed in the
constituent quark model. It is obvious that the $(\bar q\bar q)(qq)$
and $(\bar qq)(\bar qq)$ states may mix with each other. It is hard
to distinguish their internal color and flavor configurations unless
the quark dynamics is well-known or a special observable is
established. The case for crypto-exotic four-quark state is more
complicated for the mixture with normal $q\bar q$ mesons.

QCD sum rules are a powerful technique for  studying hadrons. To get
a reliable prediction, suitable currents are required. However,
there is no direct correspondence between the constituent quark
picture and the operator (current) quark picture. The diquark
concept is therefore not meaningful in the framework of a sum rule
approach.

The masses of  ``good" diquarks are obtained through an updated  sum
rule analysis. As an approximation, the resulting masses may be
identified as the masses of corresponding constituent diquark. In
terms of these constituent diquarks, the masses of light tetraquark
states are estimated. Based upon these results, it is reasonable to
identify $f_0(600)$ (or $\sigma$), $f_0(980)$, $a_0(980)$ and the
unconfirmed $\kappa(800)$ as the $0^{++}$ light tetraquark states.
However, further study of the internal quark dynamics is necessary
to determine whether diquark clusters actually exist within exotic
states.

The new observations by BES are qualitatively analyzed. The $p\bar
p$ enhancements, $X(1835)$ and $X(1812)$ are unlikely to be
tetraquark states. $X(1576)$ may be the $1^{--}$ tetraquark state
(first orbital excitations of $a_0(980)$ or $f_0(980)$) with an
``exotic" large orbital excitation. If this hypothesis is true, a
whole family of $1^{--}$ tetraquark states corresponding to orbital
excitations of $f_0(600)$ and $\kappa(800)$ may also exist, and it
will be interesting to study their radiative decays into these
scalars.


\noindent
{\bf Acknowledgments:}\\
\noindent Tao Huang is supported in part by National Natural Science
Foundation of China. T.\ Steele is supported  by NSERC (Natural
Sciences and Engineering and Research Council of Canada).


\begin{thebibliography}{50}
\bibitem{four1}
J. L. Rosner, Phys. Rev. Lett. {\bf 21}, 950 (1968).
\bibitem{four2}
R.L. Jaffe, Phys. Rev. {\bf D15}, 267 (1977); {\bf D15}, 281 (1977).
\bibitem{chan}
Hong-Mo Chan and H. Hogaasen, Phys. Lett. {\bf B72}, 121 (1977).
\bibitem{four3}
J.D. Weinstein and N. Isgur, Phys.Rev. {\bf D27}, 588 (1983).
\bibitem{mole1}
L.B. Okun and M.B. Voloskin, JETP. Lett. {\bf 23}, 333 (1976).
\bibitem{mole2}
M. Bander, G. L. Shaw, P. Thomas and S. Meshkov, Phys. Rev. Lett.
{\bf 36}, 695 (1976).
\bibitem{mole3}
A.De Rujula, H. Georgi and S.L. Glashow, Phys. Rev. Lett. {\bf 38},
317 (1977).
\bibitem{isgur2}
J.D. Weinstein and N. Isgur, Phys.Rev. {\bf D41}, 2236 (1990).
\bibitem{torn}
N.A. Tornqvist, Phys. Rev. Lett, {\bf 67}, 556 (1991).
\bibitem{frank}
F.E. Close and N.A. Tornqvist, J. Phys. {\bf G28}, R249 (2002).
\bibitem{wong}
C. Y. Wong, Phys. Rev. {\bf C69}, 055202 (2004).
\bibitem{maiani}
L. Maiani, F. Piccinini, A.D. Polosa and V. Riquer, Phys. Rev. Lett.
{\bf 93}, 212002 (2004).
\bibitem{maiani1}
L. Maiani, F. Piccinini, A.D. Polosa and V. Riquer, Phys. Rev. {\bf
D71}, 014028 (2005).

\bibitem{fariborz} D. Black, A.H. Fariborz, J. Schechter, Phys. Rev. {\bf D59},
074026 (1999).

\bibitem{others}
N.N. Achasov, S.A. Devyanin and G.N. Shestakov, Phys. Lett. {\bf
B108} (1982), 134; H.J. Lipkin, Phys. Lett. {\bf B172}, 242 (1986);
E.V. Beveren, {\it et al.}, Z. Phys. {\bf C30}, 615 (1986); N.N.
Achasov and V.N. Ivanchenko, Nucl. Phys. {\bf B315}, 465 (1989); D.
Morgan and M.R. Pennington, Phys. Rev. {\bf D48}, 1185 (1993); J.A.
Oller and E. Oset, Nucl. Phys. {\bf A620}, 438 (1997); Erratum-ibid.
{\bf A652}, 407 (1999); J.R. Pelaez, Phys. Rev. Lett. {\bf 92},
102001 (2004); I.V. Anikin, B. Pire and O.V. Teryaev, Phys. Lett.
{\bf B626}, 86 (2005).

\bibitem{pdg}
W.-M. Yao, {\it et al.}, (Particle Data Group), J. Phys. G {\bf 33},
1(2006).
\bibitem{belle}
S.K. Choi, {\it et al.}, Belle collaboration, Phys. Rev. Lett. {\bf
91}, 262001(2003).
\bibitem{babar}
B. Aubert, {\it et al.}, BaBar collaboration, Phys. Rev. Lett. {\bf
95}, 142001(2005).
\bibitem{x1}
F. E. Close and P. R. Page, Phys. Lett. {\bf B578}, 119 (2004).
\bibitem{x2}
M. B. Voloshin, Phys. Lett. {\bf B579}, 316 (2004).
\bibitem{x3}
E. S. Swanson, Phys. Lett. {\bf B588}, 189 (2004).
\bibitem{x4}
N. A. Tornqvist, Phys. Lett. {\bf B590}, 209 (2004).
\bibitem{bes1}
J.Z. Bai, {\it et al.}, BES collaboration, Phys. Rev. Lett. {\bf
91}, 022001 (2003).
\bibitem{bes2}
M. Ablikim, {\it et al.}, BES collaboration, Phys. Rev. Lett. {\bf
95}, 262001(2005).
\bibitem{bes3}
M. Ablikim, {\it et al.}, BES collaboration, Phys. Rev. Lett. {\bf
96}, 162002(2006).
\bibitem{x1576}
M. Ablikim, {\it et al.}, BES collaboration, Phys. Rev. Lett. {\bf
97}, 142002 (2006).
\bibitem{bes4}
D. P. Roy, J. Phys. {\bf G30}, R113 (2004).
\bibitem{bes5}
Bing An Li, Phys. Rev. {\bf D74}, 054017 (2006).
\bibitem{shen}
Feng-Kun Guo and Peng-Nian Shen, Phys. Rev. {\bf D74}, 097503
(2006).
\bibitem{lipkin}
M. Karliner and H.J. Lipkin, hep-ph/0607093.
\bibitem{ding}
Gui-Jun Ding and Mu-Lin Yan, Phys. Lett. {\bf B643}, 33 (2006).
\bibitem{gell}
M. Gell-Mann, Phys. Lett. {\bf 8}, 214 (1964).
\bibitem{di1}
M. Anselmino, P. Kroll and B. Pire, Z. Phys. {\bf C36}, 89 (1987).
\bibitem{di2}
H.G. Dosch, M. Jamin and B. Stech, Z. Phys. {\bf C42}, 167 (1989).
\bibitem{di3}
M. Anselmino, E. Predazzi, S. Ekelin, S. Fredriksson and D.B.
Lichtenberg, Rev. Mod. Phys. {\bf 65}, 1199 (1993).
\bibitem{di4}
R.L. Jaffe and F. Wilczek, Phys. Rev. Lett. {\bf 91}, 232003(2003).
\bibitem{di5}
M. Karliner and H.J. Lipkin, Phys. Lett. {\bf B575}, 249 (2003).
\bibitem{di6}
Shuryak and I. Zahed, Phys. Lett. {\bf B589}, 21 (2004).
\bibitem{di7}
R.L. Jaffe, Phys. Rept, {\bf 409}, 1 (2005).
\bibitem{di9}
A. Selem and F. Wilczek, hep-ph/0602128.
\bibitem{maiani2}
L. Maiani, F. Piccinini, A.D. Polosa and V. Riquer, AIP Conf. Proc.
{\bf 756}: 321 (2005).
\bibitem{sum}
M.A. Shifman, A.I. Vainshtein and V.I. Zakharov, Nucl. Phys. {\bf
B147}, 385 (1979).
\bibitem{zhang}
Ailin Zhang, Phys. Rev. {\bf D61}, 114021 (2000).
\bibitem{nielsen}
M.E. Bracco, A. Lozea, R.D. Matheus, F.S. Navarra, M. Nielsen, Phys.
Lett. {\bf B624}, 217 (2005).
\bibitem{kim}
Hungchong Kim and Yongseok Oh, Phys. Rev. {\bf D72}, 074012 (2005).
\bibitem{zhu}
Hua-Xing Chen, A. Hosaka and Shi-Lin Zhu, Phys. Rev. {\bf D74},
054001 (2006).
\bibitem{schafer} Thomas Schafer, Phys. Rev. {\bf
D68}, 114017 (2003).
\bibitem{baryon1}
B.L. Ioffe, Nucl. Phys. {\bf B188}, 317 (1981).
\bibitem{baryon2}
Y. Chung, H. G. Dosch, M. Kremer, D. Schall, Phys. Lett. {\bf B102},
175 (1981).

\bibitem{brink} D.M. Brink, Fl. Stancu, Phys. Rev. {\bf D57}, 6778 (1998).


\bibitem{melikhov}
D. Melikhov and S. Simula, Eur. Phys. J. {\bf C37}, 437 (2004).

\bibitem{tarrach}  S. Narison, R. Tarrach, Phys. Lett. {\bf B125}, 217 (1983).
\bibitem{ukawa}
S. Ohta, M. Fukugita and A. Ukawa, Phys. Lett. {\bf B173}, 15
(1986).
\bibitem{bali}
G.S. Bali, Phys. Rept. {\bf 343}, 1 (2001).
\bibitem{di10}
M. Jamin and M. Neubert, Phys. Lett. {\bf B238}, 387 (1990).
\bibitem{maiani3}
L. Maiani, F. Piccinini, A.D. Polosa and V. Riquer, Phys. Rev. {\bf
D72}, 031502(R) (2005).
\bibitem{zhang1}
Yan-Mei Kong and Ailin Zhang, hep-ph/0610245.
\bibitem{para1}
B.L. Ioffe, Prog. Part. Nucl. Phys, {\bf 56}, 232 (2006).
\bibitem{para2}
S. Narison, Phys. Lett. {\bf B626}, 101 (2005).
\bibitem{model1}
A.De. Rujula, H. Georgi and S.L. Glashow, Phys. Rev. {\bf D12}, 147
(1975).
\bibitem{tye}
W. Buchmuller and S.H.H. Tye, Phys. Rev. {\bf D24}, 132 (1981).
\bibitem{isgur}
S. Godfrey and N. Isgur, Phys. Rev. {\bf D32}, 189 (1985).
\bibitem{eichten}
E. Eichten, K. Gottfried, T. Kinoshita, K.D. Lane and T.M. Yan,
Phys. Rev. {\bf D17}, 3090(1978)[Erratum-ibid. {\bf D21}, 313
(1980)].
\bibitem{glozman}
L.Y. Glozman and D.O. Riska, Phys. Rept. {\bf 268}, 263 (1996).
\bibitem{metsch}
M. Koll, R. Ricken, D. Merten, B.C. Metsch and H.R. Petry, Eur.
Phys. J. {\bf A 9}, 73 (2000).
\bibitem{ebert}
D. Ebert, R.N. Faustov and V.O. Galkin, Phys. Lett. {\bf B634}, 214
(2006).
\bibitem{pelaez}
J.R. Pelaez, G. Rios, Phys. Rev. Lett. {\bf 97}, 242002 (2006).

\bibitem{belle2}
M.-Z. Wang {\it et al.}, Belle collaboration, Phys. Lett. {\bf
B617}, 141 (2005).

\end{thebibliography}
\end{document}